\documentclass[11pt,a4paper,reqno]{amsart}

\usepackage{times}
\usepackage{xspace}
\usepackage[inline]{enumitem}
\usepackage[numbers,sort&compress]{natbib}
\usepackage[margin=2.50cm]{geometry}
\usepackage[colorlinks=true, linkcolor=blue,citecolor=blue,urlcolor=blue]{hyperref}
\usepackage[capitalize]{cleveref}
\usepackage{tikz}
\usetikzlibrary{shapes.geometric}
\definecolor{nodeblue}{RGB}{173,216,230} 
\usetikzlibrary{positioning, arrows.meta, calc}

\usepackage{amsfonts,comment}
\usepackage{amsmath}
\usepackage{amssymb,amsmath,latexsym}

\numberwithin{equation}{section}
\numberwithin{figure}{section}
\numberwithin{table}{section}

\usepackage{color,graphicx,colortbl}
\usepackage{xcolor}

\usepackage{amsthm}
\usepackage{tabularx}   
\usepackage{booktabs}   
\usepackage{longtable}  
\usepackage{placeins}
\usepackage{graphicx}
\usepackage{multirow}
\usepackage[textwidth=20mm,textsize=small]{todonotes}
\usepackage{enumerate}
\usepackage{enumitem}

\usepackage{algpseudocode}  
\usepackage{algorithm}
\def\ue{\mathrm e}
\def\ud{\mathrm d}

\title[Shapley-based Structural Analysis of Neural Calibration]{Shapley-based Structural Analysis of Neural Calibration\\for Stochastic Volatility Models}
\author{Sha\"in Afzali}
\author{Serena Della Corte}
\author{Antonis Papapantoleon}

\address{Delft Institute of Applied Mathematics, EEMCS, TU Delft, 2628 Delft, The Netherlands}
\email{shainafzali@icloud.com}

\address{Leiden University, Mathematical Institute, 2333 CC Leiden, The Netherlands}
\email{s.della.corte@math.leidenuniv.nl}

\address{Delft Institute of Applied Mathematics, EEMCS, TU Delft, 2628 Delft, The Netherlands \& Institute of Applied and Computational Mathematics, FORTH, 70013 Heraklion, Greece}
\email{a.papapantoleon@tudelft.nl}


\begin{document} 

\begin{abstract}
%
Neural network-based approaches have emerged as efficient alternatives to traditional optimi\-zation-based procedures for the calibration of stochastic volatility models. 
However, existing work has focused primarily on predictive accuracy, with comparatively little attention devoted to understanding the structure of the learned inverse calibration mappings. 
In this work, we analyze neural calibration mappings for the Heston and rough Heston models across multilayer perceptron, highway, and softmax-parametrized highway architectures, using complementary Shapley-based methods from explainable AI. 
Specifically, we consider SHAP and $\nu$SHAP explanations, which capture distinct, complementary notions of feature relevance, corresponding to sensitivity and sufficiency of feature subsets, respectively. 
Short maturities and smile wings consistently dominate parameter inference, and the dominant attribution structure remains qualitatively stable across architectures despite differences in predictive accuracy and parameter count. 
Parameter-specific differences between SHAP and $\nu$SHAP further reveal how distinct regions of the implied volatility surface contribute to parameter recovery and expose substantial redundancy in the calibration input. 
Building on this redundancy, we show that $\nu$SHAP explanations can guide a significant reduction in input dimensionality for the rough Heston model while matching calibration accuracy relative to the full implied volatility surface. 
These findings demonstrate that complementary Shapley-based methods provide structural insight into learned inverse calibration mappings beyond predictive error metrics, and offer a practical route to feature selection in neural calibration problems.
\end{abstract}

\maketitle


\section{Introduction}

The last few years have been marked by an increasing interest in neural network-based approaches for the calibration of stochastic volatility models, offering efficient alternatives to traditional optimization-based procedures. 
By learning the inverse mapping from implied volatility surfaces to model parameters, such methods replace repeated optimization steps with offline training, enabling near-instantaneous inference during deployment. 
A growing body of literature has demonstrated that neural networks are capable of achieving accurate calibration for both classical and rough stochastic volatility models; see \textit{e.g.} \citet{hernandez_2016,roder_dimitroff_2020,horvath_2021,risks8040101,liu2019neural,baschetti2024deep,bayer2025on} and \citet{sridi_bilokon_2023}.

Despite these advances, existing work has focused primarily on predictive accuracy, while comparatively little attention has been devoted to understanding the structure of the learned inverse mappings themselves. 
Inverse calibration may be weakly identifiable or ill-conditioned, since different parameter configurations may generate highly similar implied volatility surfaces; see \textit{e.g.} \citet{HeinHofmann2003}. 
Consequently, predictive performance alone provides limited insight into whether the learned mapping relies on regions of the implied volatility surface that are consistent with the known qualitative effects of the model parameters. 
Recent work has investigated interpretability methods for neural calibration of (rough) stochastic volatility models, demonstrating that feature attribution methods can reveal meaningful relationships between regions of the implied volatility surface and model parameters; see \citet{yuan_brigo_jacquier_pede_2024} and \citet{brigo_huang_pallavicini_haitz_2025}. 
These articles mainly used sensitivity-based feature attribution to highlight the relevance of short maturities and smile-wing regions in the calibration process. 
While these results provide valuable insight into the behavior of neural calibration models, they leave open whether the identified regions are also sufficient to preserve parameter predictions and how feature redundancy affects the learned inverse mapping.
Establishing which regions of the surface a calibration network actually relies on is not merely of academic interest: a network that performs well on clean, synthetically generated surfaces may still depend on regions that are noisy, illiquid, or missing altogether from real market data, hence structural validation is a prerequisite for deploying such methods with confidence.

Building on these articles, we investigate the structure of neural calibration mappings for the Heston and rough Heston models using complementary Shapley-based methods from the field of explainable Artificial Intelligence (XAI). 
More specifically, we study the {SHAP} and the {$\nu$SHAP} methodologies, which provide complementary notions of feature relevance corresponding to sensitivity and sufficiency, respectively; see \citet{lundberg_lee_2017} and \citet{marquessilva_huang_letoffe_2025}.
We analyze different neural network architectures, including multilayer perceptrons, highway networks, and softmax-parametrized highway networks. 
Although the best-performing highway-based architectures yield lower prediction errors in several experiments, our results indicate that multilayer perceptrons already achieve competitive calibration accuracy with substantially fewer parameters.
The empirical analysis further reveals stable attribution structures across architectures and Shapley-based methods, with short maturities and smile wings consistently playing a dominant role in parameter inference. 
Attribution structures differ substantially across parameters, revealing parameter-specific feature relevance and redundancy within the implied volatility surfaces. 
Furthermore, we demonstrate that $\nu$SHAP explanations can be used operationally to guide input dimensionality reduction while largely preserving calibration accuracy.

The main contributions of this work are threefold: 
\begin{itemize}
    \item Introducing a complementary SHAP--$\nu$SHAP framework for the structural analysis of neural-network calibration in the context of stochastic volatility models.

    \item Providing empirical insight into parameter identifiability and redundancy within the Heston and rough Heston calibration problems, highlighting the importance of short maturities and smile-wing regions for parameter recovery.

    \item Demonstrating that $\nu$SHAP explanations can guide input dimensionality reduction while preserving comparable predictive performance using substantially fewer input features.
\end{itemize}

The remainder of this paper is organized as follows. 
Section \ref{sec:framework} introduces the calibration framework and neural network architectures, while Section \ref{sec:XAI} presents the Shapley-based methods and discusses their complementary interpretation. 
Section \ref{sec:cal_performance} describes the experimental setup and calibration performance. 
Section \ref{sec:results} presents the structural validation results including attribution analysis, identifiability, architectural robustness, and Shapley-guided input dimensionality reduction. 
Finally, Section \ref{sec:conclusion} concludes and outlines potential directions for future research.


\section{Calibration framework}
\label{sec:framework}

\subsection{Implied volatility and the calibration problem}

For a European call option with strike $K$ and maturity $T$, the implied volatility is defined as the volatility parameter for which the corresponding \citet{black_scholes_1973} price matches the observed option price; see \textit{e.g.} \citet[Ch. 1]{Gatheral_2006}. 
Evaluating implied volatilities across strikes and maturities yields an implied volatility surface. 
Model calibration is the process of identifying the parameters of a model by matching the option prices or implied volatilities produced by the model to those observed in the market.
In the calibration setting, market information is represented by a finite collection of option contracts $\{\zeta_i\}_{i=1}^N$, corresponding to a discretization of the implied volatility surface across strikes, or moneyness levels, and maturities.

We follow the calibration formulation as defined by \citet{horvath_2021}. 
Let $M := M(\mathfrak{p})_{\mathfrak{p} \in \mathcal P}$ denote a model with parameters $\mathfrak{p}$ in the parameter set $\mathcal P \subset \mathbb{R}^n$. 
Consider a financial contract $\zeta$, such as a European call option characterized by strike and maturity, and let $P$ denote the pricing operator mapping the model and contract to a model-implied quantity. 
In our setting, this quantity corresponds to the implied volatility. Thus,
$$
P(M(\mathfrak{p}),\zeta)
$$
denotes the model-implied volatility of contract $\zeta$ under parameters $\mathfrak{p}$.

Let $P^{\mathrm{mkt}}(\zeta)$ denote the implied volatility observed in the market. 
Given the finite collection of contracts $\{\zeta_i\}_{i=1}^N$, model calibration consists of determining parameters $\mathfrak{p}$ such that the model-implied quantities match the observed market data as closely as possible. Given a suitable metric $\delta(\cdot,\cdot)$, this leads to the optimization problem
\begin{equation}
    \mathfrak{p}^* \in \arg\min_{\mathfrak{p} \in \mathcal P}
    \delta\left(
    \{P(M(\mathfrak{p}),\zeta_i)\}_{i=1}^N,
    \{P^{\mathrm{mkt}}(\zeta_i)\}_{i=1}^N
    \right).
\end{equation}
Instead of viewing calibration purely as an optimization problem, it is useful to interpret it as an inverse problem. The forward map
\begin{equation}
    \mathfrak{p} \mapsto \{P(M(\mathfrak{p}),\zeta_i)\}_{i=1}^N
\end{equation}
maps model parameters to observable quantities, namely implied volatilities. 
Calibration seeks to invert this mapping, \textit{i.e.} to construct a function
\begin{equation}
    \Theta : \mathbb{R}^N \to \mathcal P
    \quad\text{ such that } \quad
    \{P^{\mathrm{mkt}}(\zeta_i)\}_{i=1}^N \mapsto \mathfrak{p}.
\end{equation}

In this work, we approximate the inverse map $\Theta$ directly using neural networks, as proposed by \citet{hernandez_2016}, \citet{roder_dimitroff_2020}, \citet{horvath_2021} and \citet{bayer2025on}.
In order to construct training data for supervised learning, parameter vectors are sampled from bounded parameter domains and subsequently mapped to implied volatility surfaces through the forward pricing operator. 
The resulting surfaces are discretized across a moneyness--maturity grid, yielding fixed-dimensional feature vectors which serve as inputs to the neural networks.

Since observed market implied volatility surfaces are not accompanied by known ground-truth model parameters, we use synthetically generated surfaces to construct the supervised learning dataset. 
In order to obtain broad coverage of the parameter space, we employ Latin Hypercube Sampling (LHS; \citet{mckay_beckman_connover_1979}), which generates stratified samples across the admissible parameter domain while avoiding excessive clustering in high-dimensional regions.
Once trained offline on synthetically generated implied volatility surfaces with known parameters, the calibration step reduces to a single forward pass through the network.


\subsection{Heston and rough Heston models}

The Heston model \cite{heston_1993} is among the most widely used stochastic volatility models in quantitative finance. 
The asset price and the variance evolve stochastically under the risk-neutral measure $\mathbb{Q}$. 
The dynamics are provided by
\begin{align} \label{eq:asset_dynamics}
    \ud S_t &= rS_t \, \ud t + \sqrt{v_t}S_t \, \ud W_t^S \\
    dv_t &= \kappa(\bar v - v_t) \, \ud t + \gamma\sqrt{v_t} \, \ud W_t^v,
\end{align}
with $\kappa, \gamma, v_0, \bar{v} > 0$. 
Here, $S$ denotes the asset price process and $v$ the instantaneous variance process. 
Both processes are driven by standard Brownian motions $W^v$ and $W^S$ with correlation $\rho \in [-1,1]$. 
Thanks to the affine nature of the Heston model, we can compute option prices efficiently using the COS method of \citet{fang_oosterlee_2009} or Fourier transform methods; see \textit{e.g.} \citet{fourier_transform_valuation}. 
Implied volatilities are subsequently extracted using J\"ackel's algorithm \cite{Jackel_2015}.

Although the Heston model successfully captures important stylized features such as volatility smiles and skews, empirical evidence presented by \citet{gatheral_jaisson_rosenbaum_2018} suggests that volatility trajectories exhibit substantially rougher behavior than predicted by classical diffusion-based models. 
In particular, realized volatility exhibits scaling properties consistent with fractional dynamics possessing Hurst parameter $H \ll \frac12$, typically around $H \approx 0.1$; see \cite{gatheral_jaisson_rosenbaum_2018}.

In order to account for this phenomenon, rough volatility models introduce memory into the variance dynamics through Volterra-type processes. 
A general stochastic Volterra process may be written as
\begin{equation}
    v_t
    =
    v_0
    +
    \int_0^t K(t-s) \, b(v_s) \, \ud s
    +
    \int_0^t K(t-s) \, \sigma(v_s) \, \ud W_s,
\end{equation}
where $K$ denotes a deterministic kernel function controlling the influence of past states on the current dynamics; see \emph{e.g.} \citet{jaber_larsson_pulido_2019}.
A commonly used choice, motivated by fractional models, is the fractional kernel
\begin{equation} \label{eq:rough_kernel}
    K(t)
    =
    \frac{t^{H-\frac12}}
    {\Gamma\left(H+\frac12\right)},
\end{equation}
where $H \in \left(0,\frac12\right)$ denotes the Hurst parameter and $\Gamma(\cdot)$ the gamma function. 
Since $H < \frac12$, the kernel becomes singular near the origin, inducing rough local behavior.

We consider the rough Heston model introduced by \citet{el_euch_fukusawa_rosenbaum_2018}. 
In this model, the asset price dynamics are still provided by (\ref{eq:asset_dynamics}), while the variance process satisfies
\begin{equation}
    v_t
    =
    v_0
    +
    \int_0^t
    K(t-s) \, \kappa(\bar v - v_s) \, \ud s
    +
    \int_0^t
    K(t-s) \, \gamma\sqrt{v_s} \, \ud W_s^v,
\end{equation}
with fractional kernel $K$ defined by (\ref{eq:rough_kernel}), and correlation structure as in the classical Heston model. 
Unlike the classical Heston model, the rough Heston variance process is not Markovian due to its dependence on the full historical trajectory through the kernel $K$.

European option prices under the rough Heston model can be computed using the characteristic function representation derived by \citet{el_euch_rosenbaum_2018}. 
In this work, we solve the corresponding fractional Riccati equation numerically using the fractional Adams method as presented by \citet{diethelm_ford_freed_2004}, after which option prices are obtained through the COS method. 
Similarly to the Heston case, we compute implied volatilities using J\"ackel's algorithm.


\subsection{Neural network architectures} \label{sect:architectures}

In order to approximate the inverse calibration map, we consider several feedforward neural network architectures of increasing structural complexity, namely multilayer perceptrons (MLPs), highway networks, and softmax-parametrized highway networks.

The multilayer perceptron serves as a flexible baseline architecture consisting of fully connected layers interleaved with non-linear activation functions. 
Despite their relative simplicity, MLPs have demonstrated strong performance in inverse calibration problems for stochastic volatility models; see \textit{e.g.} \citet{brigo_huang_pallavicini_haitz_2025, bayer2025on, horvath_2021} and \citet{yuan_brigo_jacquier_pede_2024}.

\begin{figure}[htbp]
    \centering
    \resizebox{0.7\textwidth}{!}{\begin{tikzpicture}[
    >=Stealth,
    node distance=2cm,
    input/.style={circle, draw=cyan!80, fill=cyan!30, thick, minimum size=1.2cm},
    block/.style={rectangle, draw=orange!60, fill=orange!30, thick, rounded corners, minimum width=3.8cm, minimum height=1cm, align=center},
    state/.style={circle, draw=green!80, fill=green!40, thick, minimum size=1.2cm},
    output/.style={circle, draw=purple!60, fill=purple!30, thick, minimum size=1.2cm}
]

\node[input] (x) at (0, 0) {$x$};

\node[block] (Wt) at (4.5, 0) {$W_T x + b_T$};
\node[block] (Wh) at (4.5, 2) {$W_H x + b_H$};

\node[state] (T) at (8.5, 0) {$T$};
\node[state] (H) at (8.5, 2) {$H$};

\node[block, minimum width=4.5cm] (Combine) at (13, 0) {$H \odot T + x \odot (1 - T)$};

\node[output] (y) at (17, 0) {$y$};

\draw (x.east) -- (2, 0);
\draw[->] (2, 0) |- (Wh.west);
\draw[->] (2, 0) -- (Wt.west);

\draw[->] (Wh.east) -- node[above] {$g_H$} (H.west);
\draw[->] (Wt.east) -- node[above] {$\sigma$} (T.west);

\draw[->] (H.east) -- ([yshift=0.2cm]Combine.west);
\draw[->] (T.east) -- (Combine.west);

\draw[->] (x.south) -- +(0, -0.5) -| (Combine.south);

\draw[->] (Combine.east) -- (y.west);

\end{tikzpicture}} 
    \caption{Schematic representation of a single highway layer, \emph{cf.} (\ref{eq:highway}). }
    \label{fig:highway_layer}
\end{figure}
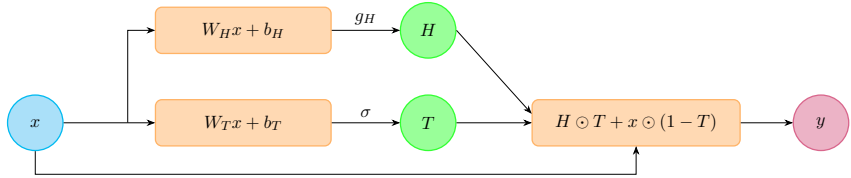

In an effort to facilitate the information flow in deeper architectures, we additionally consider highway networks introduced by \citet{srivastava_greff_schmidhuber_2015}.
Highway networks augment standard feedforward layers with learnable gating mechanisms that regulate the propagation of transformed and carried information across layers.
Let
\begin{equation}
    H = g_H(W_Hx + b_H)
\end{equation}
denote a non-linear transformation of the input $x$, and let
\begin{equation}
    T = \sigma(W_Tx + b_T)
\end{equation}
denote a transform gate, where $\sigma$ denotes the sigmoid activation function. 
A highway layer is then defined by
\begin{equation} \label{eq:highway}
    y = H \odot T + x \odot (1-T),
\end{equation}
where $\odot$ denotes the Hadamard product (\textit{i.e.} element-wise multiplication). 
A visual representation of a single highway layer is shown in Figure~\ref{fig:highway_layer}.

\noindent The gate $T$ determines component-wise how much transformed information is propagated through the layer and how much of the original input is preserved. 
Consequently, the layer output is a convex combination of transformed and propagated information. This controlled information flow helps stabilize optimization and mitigate degradation effects in deeper architectures.

In addition to classical highway networks, we consider a two-gate extension in which the transform and carry branches are parametrized separately. Introducing the pre-activations
$$
z_T = W_Tx + b_T,
\qquad
z_C = W_Cx + b_C,
$$
the layer output is written as
\begin{equation} \label{eq:gen_highway}
y = H \odot T + x \odot C,
\end{equation}
where $T$ and $C$ denote the transform and carry gates, respectively.
In the unconstrained formulation, the gates are computed independently,
$$
T = g_T(z_T),
\qquad
C = g_C(z_C),
$$
so that, in general,
$$
T + C \neq 1.
$$
This formulation, usually called a \textit{generalized} highway network, is genuinely more flexible than the standard highway layer, since the transform and carry contributions are not restricted to be complementary. However, this additional flexibility can lead to unstable signal propagation in deep networks. 
If both gates become simultaneously large, the hidden representation may be amplified from layer to layer, whereas small values of both gates may lead to excessive attenuation. 
In deep configurations, such deviations can accumulate across layers and result in exploding or vanishing hidden representations.

Therefore, in order to stabilize the architecture, we impose a joint softmax normalization on the transform and carry pre-activations:
$$
(T,C)=\operatorname{softmax}(z_T,z_C),
$$
\textit{i.e.} component-wise we have
\begin{equation} \label{eq:softmax_gates}
T_i =
\frac{\ue^{(z_T)_i}}
{\ue^{(z_T)_i}+\ue^{(z_C)_i}},
\qquad
C_i =
\frac{\ue^{(z_C)_i}}
{\ue^{(z_T)_i}+\ue^{(z_C)_i}}.
\end{equation}
Hence, by construction,
$
T + C = 1
$
holds element-wise, so that the layer output is again a convex combination of the transformed and carried representations.

\begin{figure}[htbp]
    \centering
    \resizebox{0.7\textwidth}{!}{\begin{tikzpicture}[
    >=Stealth,
    node distance=2cm,
    input/.style={circle, draw=cyan!80, fill=cyan!30, thick, minimum size=1.2cm},
    block/.style={rectangle, draw=orange!60, fill=orange!30, thick, rounded corners, minimum width=3.8cm, minimum height=1cm, align=center},
    block_softmax/.style={rectangle, draw=orange!60, fill=orange!70, thick, rounded corners, minimum width=3.8cm, minimum height=1cm, align=center},
    state/.style={circle, draw=green!80, fill=green!40, thick, minimum size=1.2cm},
    output/.style={circle, draw=purple!60, fill=purple!30, thick, minimum size=1.2cm}
]

\node[input] (x) at (0, 0) {$x$};

\node[block] (Wh) at (4.5, 2) {$W_H x + b_H$};
\node[block] (Wt) at (4.5, 0) {$W_T x + b_T$};
\node[block] (Wc) at (4.5, -2) {$W_C x + b_C$};

\node[block_softmax, minimum height=1cm, minimum width=2cm] (Softmax) at (8.5, -1) {softmax};

\node[state] (H) at (11.5, 2) {$H$};
\node[state] (T) at (11.5, 0) {$T$};
\node[state] (C) at (11.5, -2) {$C$};

\node[block, minimum width=4.5cm] (Combine) at (16, 0) {$H \odot T + x \odot C$};

\node[output] (y) at (20, 0) {$y$};

\draw (x.east) -- (2.2, 0);
\draw[->] (2.2, 0) |- (Wh.west);
\draw[->] (2.2, 0) -- (Wt.west);
\draw[->] (2.2, 0) |- (Wc.west);

\draw[->] (Wh.east) -- node[above] {$g_H$} (H.west);

\draw[->] (Wt.east) -- ([yshift=0.4cm]Softmax.west);
\draw[->] (Wc.east) -- ([yshift=-0.4cm]Softmax.west);

\draw[->] ([yshift=0.4cm]Softmax.east) -- (T.west);
\draw[->] ([yshift=-0.4cm]Softmax.east) -- (C.west);

\draw[->] (H.east) -- ([yshift=0.4cm]Combine.west);
\draw[->] (T.east) -- (Combine.west);
\draw[->] (C.east) -- ([yshift=-0.4cm]Combine.west);

\draw[->] (x.south) -- +(0, -2.5) -| (Combine.south);

\draw[->] (Combine.east) -- (y.west);

\end{tikzpicture}} 
    \caption{Schematic representation of a single softmax-parametrized highway layer, \emph{cf.} (\ref{eq:gen_highway}), (\ref{eq:softmax_gates}). }
    \label{fig:gen_highway_layer}
\end{figure}
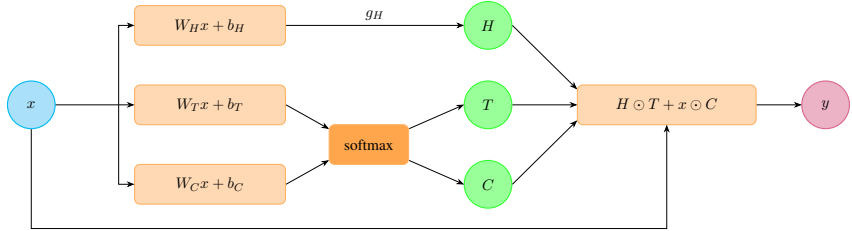

An important observation is that the softmax constraint removes the additional functional freedom of the unconstrained two-gate formulation. 
Indeed, for a two-way softmax,
$$
T = \sigma(z_T-z_C)
\quad \text{ and } \quad
C = 1-T.
$$
Substituting the gate pre-activations yields
$$
T
=
\sigma\left((W_T-W_C)x + (b_T-b_C)\right),
$$
which is equivalent to the gating function of a standard highway layer with effective parameters
$$
W_{\mathrm{eff}} = W_T-W_C
\quad \text{ and } \quad
b_{\mathrm{eff}} = b_T-b_C.
$$
Hence, the constrained architecture does not define a larger function class than the classical Highway network. 
The distinction lies instead in its two-logit, overparametrized gate formulation, in which transform and carry pre-activations are learned separately and compete through the softmax normalization, thereby inducing different parametrization and optimization dynamics despite an equivalent function class. 
A visual representation of this softmax-parametrized highway layer is shown in Figure~\ref{fig:gen_highway_layer}.

\section{Shapley-based framework for Structural Analysis} 
\label{sec:XAI}

We will employ Shapley-based methods from explainable AI to analyze the structure of the learned inverse calibration map. 
The aim is not merely to explain individual predictions, but to assess whether the network relies on regions of the implied volatility surface that are consistent with the known qualitative effects of the model parameters. 
We therefore interpret structural analysis as the assessment of whether the learned inverse map relies on model-consistent information and exhibits stable, interpretable attribution patterns. 
In our context, an attribution pattern is considered structurally meaningful if it consistently highlights regions of the surface that are known to contain information about the corresponding parameter, remains relatively stable across architectures, and admits a consistent interpretation across complementary Shapley methods. 
To this end, we consider two attribution methods: SHAP, see \citet{lundberg_lee_2017}, and $\nu$SHAP, see \citet{marquessilva_huang_letoffe_2025}. 
These methods are based on related game-theoretic principles, but differ in the value function used to assign importance to feature coalitions. As a result, they capture different notions of feature relevance.
The Shapley-based methods sit alongside other model-agnostic and gradient-based attribution approaches developed in the explainable AI literature, such as LIME and integrated gradients; see \citet{ribeiro2016why} and \citet{sundararajan2017axiomatic}.
We focus on Shapley-based methods for their axiomatic grounding and their natural sensitivity and sufficiency duality, discussed in subsection \ref{sec:complementary}.


\subsection{Explanation framework}
\label{subsec:XAI-frame}

Let us introduce the notation following the logic-based explainability framework for regression models described by \citet{marquessilva_2025}. 
In our application of this framework, explanations are constructed locally for individual instances and subsequently aggregated to analyze global attribution structure.

Let $\mathcal{I} = \{1,\ldots,M\}$ denote the set of input features. Each feature $i \in \mathcal{I}$ takes values in a domain $\mathbb{D}_i$, and the feature space is given by $\mathbb{F} = \mathbb{D}_1 \times \cdots \times \mathbb{D}_M.$
An input point is denoted by 
$$
x = (x_1,\ldots,x_M) \in \mathbb{F},
$$ 
while 
$$
v = (v_1,\ldots,v_M) \in \mathbb{F}
$$
denotes the input point to be explained. In the present setting, $v$ corresponds to a discretized implied volatility surface.

A regression model is characterized by a non-constant function
\begin{equation}
    \xi : \mathbb{F} \to \mathbb{V},
\end{equation}
where $\mathbb{V}\subseteq \mathbb{R}$ denotes the output space. In our calibration setting, the neural network output is a vector of model parameters.
Since the feature attributions are computed separately for each component, we let $\xi$ denote the scalar-valued prediction function associated with one fixed model parameter.
We represent the regression model by the tuple $\mathcal{M}_R = (\mathcal{I},\mathbb{F},\mathbb{V},\xi).$
An \emph{instance} is given by the pair $(v,q)$, where $q=\xi(v)$. Together, the regression model and the instance define an explanation problem $\mathcal{E} = (\mathcal{M}_R,(v,q)).$

Consider a feature $i \in \mathcal{I}$ and let $\approx_i$ denote a similarity relation on its domain $\mathbb{D}_i$. 
Thus, $x_i \approx_i v_i$ means that the values $x_i$ and $v_i$ are considered sufficiently close to each other. For a subset of features $S \subseteq \mathcal{I}$, we say that $x$ and $v$ are similar on $S$ if they are similar in every coordinate belonging to $S$, \textit{i.e.}
\begin{equation}
    x_S \approx v_S
    \quad \Longleftrightarrow\quad
    x_i \approx_i v_i \quad \text{for all $i \in S$.}
\end{equation}
In the special case $S=\mathcal{I}$, we simply write $x \approx v$. The specific similarity relations used in the numerical experiments are
defined in Section \ref{subsec:setup}.
An analogous similarity relation is considered in the output space $\mathbb{V}$, such that $\xi(x) \approx \xi(v)$ holds whenever the model outputs are deemed sufficiently close.

We focus on local explanation methods designed to explain the output $\xi(v)$ for a single instance $v$. 
Many local explanation models operate on simplified binary inputs $z' \in \{0,1\}^M$, which encode whether features are present or absent. 
A reconstruction map $h_v : \{0,1\}^M \to \mathbb{F}$ maps such simplified inputs back to the original feature space.
A broad class of local explanation methods is given by additive feature attribution methods, for which the explanation model takes the form
\begin{equation}
    g(z') = \psi_0 + \sum_{i=1}^M \psi_i z_i',
\end{equation}
where $\psi_i \in \mathbb{R}$ denotes the attribution assigned to feature $i$.
These local attributions can be aggregated across a collection of instances. 
Given local attributions $\psi_i(x)$ for feature $i$ and a collection of $N$ instances $\{x^{(n)}\}^N_{n=1}$, a global importance score can be obtained as
\begin{equation}
    I_i = \frac{1}{N}\sum_{n=1}^N\left|\psi_i\big(x^{(n)}\big)\right|.
\end{equation}
Equivalently, this quantity may be interpreted as the expectation of $|\psi_i(x)|$ with respect to the empirical distribution that assigns equal mass to each explained instance. The absolute value is used to measure the magnitude of feature contribution independently of sign. 

The resulting global attribution scores are arranged according to the dimensions of the original money\-ness--maturity grid and visualized as heatmaps. The Shapley-based methods are applied to scaled and ZCA-whitened inputs. Consequently, the attributed coordinates correspond to transformed features rather than directly to individual option contracts: each whitened coordinate is generally a linear combination of the original implied volatility features.  The heatmaps should therefore not be interpreted pointwise as importance scores for individual strike--maturity contracts. Instead, we use them as approximate visual summaries of broader attribution patterns in the transformed representation. This limitation should be kept in mind when relating the observed patterns to regions of the original implied volatility surface.


\subsection{SHAP}

SHAP explanations are based on the Shapley value framework from cooperative game theory, \textit{cf.} \citet{Shapley_1953}, where the input features play the role of players and a \emph{value function} assigns a payoff to each coalition of features.
Let
\begin{equation}
    \nu(\cdot,\,\mathcal{E}):2^{\mathcal{I}} \to \mathbb{R}
\end{equation}
denote a value function associated with the explanation problem $\mathcal{E}$. The attribution assigned to feature $i$ is given by the Shapley value
\begin{equation} \label{eq:shapley_value}
    \psi_i(v)
    =
    \sum_{S \subseteq \mathcal{I} \setminus \{i\}}
    \frac{|S|!\,(|\mathcal{I}|-|S|-1)!}{|\mathcal{I}|!}
    \Big\{
    \nu(S \cup \{i\},\,\mathcal{E})
    -
    \nu(S,\,\mathcal{E})
    \Big\}.
\end{equation}
This quantity measures the average marginal contribution of feature $i$ over all coalitions that do not contain $i$.
%
%
%
The Shapley value framework itself is independent of the particular choice of value function. 
Different explanation methods therefore correspond to different definitions of $\nu$.

In our implementation of SHAP, missing features are replaced by values drawn from the empirical background distribution
Let $Z$ denote a random observation from this distribution and let $\bar S=\mathcal I\setminus S$. 
We consider the background-replacement value function
\begin{equation}
    \nu_e(S,\mathcal E)
    =
    \mathbb{E}_{Z}\left[
        \xi\left(v_S,Z_{\bar S}\right)
    \right].
    \label{eq:shap_value}
\end{equation}
In particular,
\begin{equation*}
    \nu_e(\emptyset,\mathcal E)
    =
    \mathbb{E}_{Z}[\xi(Z)]
    \quad \text{ and } \quad
    \nu_e(\mathcal I,\mathcal E)
    =
    \xi(v).
\end{equation*}
The value function $\nu_e$ therefore measures how the model prediction changes when the features in $S$ are fixed at their values in the explained instance, while the remaining features are supplied by the background distribution. 
SHAP consequently provides a \emph{sensitivity-oriented} notion of feature relevance under this background-replacement scheme.

Exact computation of the Shapley values is infeasible for high-dimensional input spaces, since it requires evaluating all possible feature coalitions. 
We therefore employ KernelSHAP, see \citet{lundberg_lee_2017}, which approximates the Shapley values by fitting a weighted local linear model to model evaluations obtained from sampled coalitions.


\subsection{\texorpdfstring{$\nu$}{nu}SHAP}

While SHAP attributes importance through marginal changes in expected model output, $\nu$SHAP adopts a \emph{sufficiency-based} perspective grounded in logic-based explainability. 
The method was introduced by \citet{marquessilva_huang_letoffe_2025} as an alternative to sensitivity-oriented explanations, motivated by the observation that such explanations may be misleading in the presence of correlated or redundant features.

In the present work, however, we do \emph{not} treat SHAP and $\nu$SHAP as competing explanation methods. 
Instead, we interpret them as complementary tools. SHAP measures how strongly the model output changes when features are conditioned upon, whereas $\nu$SHAP measures whether subsets of features are sufficient to preserve the prediction. 
These are different notions of explanation and for inverse calibration problems both are informative.

To define $\nu$SHAP, we first introduce \emph{weak abductive explanations}. A subset of features $S \subseteq \mathcal{I}$ is called a weak abductive explanation ($\operatorname{WAXp}$) if keeping the features in $S$ close to their values in the explained instance guarantees that the model prediction remains close to the original prediction. 
Formally, $S\subseteq \mathcal{I}$ is a $\operatorname{WAXp}$ if $S \in \mathcal{W(E)}$, where
\begin{equation}
    \mathcal{W(E)}
    := \left\{ S \subseteq \mathcal{I} \, : \, 
    \forall x \in \mathbb{F},
    x_S \approx v_S
    \Longrightarrow
    \xi(x) \approx \xi(v)\right\}.
\end{equation}
A subset-minimal WAXp -- in the sense that no subset can be removed further -- is called an \emph{abductive explanation} ($\operatorname{AXp}$).
%
%
This definition leads to a binary value function
\begin{equation} \label{eq:nushap_valuefunction}
    \nu_s(S,\,\mathcal{E})
    =
    \begin{cases}
    1, & \text{if } S\in \mathcal{W(E)},\\
    0, & \text{otherwise}.
    \end{cases}
\end{equation}
The value function $\nu_s$ measures whether a subset of features is sufficient to preserve the prediction under the prescribed similarity relation. 

An important consequence of (\ref{eq:nushap_valuefunction}) is that $\nu$SHAP is not an additive feature attribution method in the same sense as SHAP. In particular, the attributions do not sum to the model output $\xi(v)$, but to
\begin{equation*}
    \nu_s(\mathcal{I},\,\mathcal{E})-\nu_s(\emptyset,\,\mathcal{E})=1.
\end{equation*}
Moreover, $\nu$SHAP attributions are non-negative and take values in $[0,1]$, reflecting the monotonicity of the sufficiency value function: if a subset of features is sufficient, then any superset is also sufficient.

In practice, exact verification of WAXps over the full feature space is infeasible. 
We therefore use a sample-based approximation, as suggested in \citet{marquessilva_huang_letoffe_2025}. 
Given a background dataset $D \subset \mathbb{F}$, a set $W \subseteq \mathcal{I}$ is called a \emph{sample-based weak abductive explanation} ($\operatorname{sbWAXp}$) if $W\in \mathcal{W(E)}_D$, where
\begin{equation}
\label{eq:sbwaxp}
    \mathcal{W(E)}_D := \left\{W\subseteq \mathcal{I}\, : \, 
    \forall x \in D, 
    x_W \approx v_W
    \Longrightarrow
    \xi(x) \approx \xi(v)\right\}.
\end{equation}
This condition may be verified by searching for a counterexample in the background dataset: if there exists an $x \in D$ such that $\xi(x) \not\approx \xi(v)$ while $x_W \approx v_W$, then $W$ is not an sbWAXp.

The resulting $\nu$SHAP values are approximated using a permutation-sampling estimator. For a uniformly sampled permutation $\pi$ of the feature set $\mathcal{I}$, let $S_\pi(i)$ denote the set of features preceding $i$ in the permutation. Then the $\nu$SHAP value can be written as
\begin{equation} \label{eq:nushap_perm}
    \psi_i
    =
    \mathbb{E}_{\pi}
    \left[
    \nu_s(S_\pi(i)\cup\{i\}, \, \mathcal{E})
    -
    \nu_s(S_\pi(i), \, \mathcal{E})
    \right],
\end{equation}
which is approximated via Monte Carlo sampling over permutations.

\subsection{Complementary interpretation}
\label{sec:complementary}

The distinction between SHAP and $\nu$SHAP is central to our interpretation of the results. 
SHAP measures the marginal change in the coalition value induced by conditioning on a feature and therefore provides a \emph{sensitivity}-oriented notion of relevance. 
By contrast, $\nu$SHAP evaluates whether feature subsets are \emph{sufficient} to preserve the prediction under the prescribed similarity relations. 
The two methods share the Shapley axiomatization and differ only in the underlying value function, so that any divergence between the resulting attributions is attributable to the distinct notions of relevance encoded by $\nu_e$ and $\nu_s$ rather than to differences in the aggregation scheme.

An additional distinction concerns the inputs on which the two value functions are evaluated. 
The SHAP background-replacement procedure may generate recombined inputs that do not correspond to admissible implied volatility surfaces, particularly in the presence of dependencies across strikes and maturities. 
By contrast, the sample-based sufficiency criterion used for $\nu$SHAP is evaluated only against observations contained in the background dataset. 
This difference should therefore also be taken into account when interpreting discrepancies between the two attribution structures. 
Discussion of interventional and conditional Shapley values under feature dependence appear, for example, in \citet{janzing2020,chen2020,sundararajan2020}, and \citet{aas2021}.

The two methods therefore probe the model in different regions of the input space.
The interventional value function \eqref{eq:shap_value} evaluates the network on recombined inputs which need not resemble any admissible implied volatility surface, whereas the sample-based sufficiency criterion \eqref{eq:sbwaxp} is verified only against instances contained in the background dataset and hence remains confined to the support of the training distribution. 
This asymmetry constitutes a second source of divergence between the two attribution structures, in addition to the sensitivity--sufficiency distinction itself, and we take it into account when interpreting the empirical results.

The distinction is particularly relevant for implied volatility surfaces, which exhibit strong dependencies across neighboring strikes and maturities. 
Multiple regions of the surface may therefore carry overlapping information about the same model parameter, in which case sensitivity and sufficiency need not coincide: a feature may substantially alter the model output when intervened upon, while a correlated region elsewhere on the surface still retains enough information to preserve the prediction.

This is the reason why disagreement between SHAP and $\nu$SHAP should not be read as a failure of either method. 
Agreement indicates that a region is both influential and sufficient for parameter recovery, and thus that the corresponding information is localized. 
Divergence admits several readings, which the analysis of Section~\ref{sec:results} seeks to separate: redundancy within the implied volatility surface, interactions between parameters, weak identifiability of the inverse map, or sensitivity of the network to off-distribution inputs. 
The interpretations we draw below are accordingly stated as qualitative evidence rather than as identification results.

In the empirical analysis that follows, we use both methods to investigate parameter-specific attribution structures, architectural robustness, redundancy within implied volatility surfaces, and the extent to which these attribution patterns support meaningful parameter identifiability.


\subsection{Experimental explanation setup}\label{subsec:setup}

Local SHAP and $\nu$SHAP explanations are computed for $50$ randomly sampled test instances, for each trained neural network. 
The resulting local explanations are subsequently aggregated in order to analyze the global attribution structure. 
Global SHAP heatmaps are obtained by averaging the absolute SHAP values across explained instances, while global $\nu$SHAP heatmaps are obtained by averaging the corresponding $\nu$SHAP values. 
The resulting attribution scores are normalized and visualized on the original moneyness--maturity grid.

We use a background dataset $\mathcal B \subset \mathbb{F}$ consisting of $100$ randomly sampled validation instances for both explanation methods. 
This dataset represents the empirical background distribution used to evaluate coalition values for KernelSHAP. 
The same dataset is used to search for counterexamples to the sufficiency condition for the sample-based $\nu$SHAP approximation. 

The input-similarity relation is evaluated in the scaled and ZCA-whitened feature space. 
Consider a subset $S\subseteq\mathcal{I}$, then we define 
\begin{equation}
    x_S\approx v_S
    \quad\Longleftrightarrow\quad
    |x_i-v_i|\leq\varepsilon_{\mathrm{in}}
    \quad\text{for every }i\in S,
\end{equation}
where $\varepsilon_{\mathrm{in}}=0.05.$
Output similarity is defined separately for each calibrated parameter.
Let $\xi_j$ denote the prediction function associated to the output component $j$. 
Then, we define
\begin{equation}
    \xi_j(x)\approx\xi_j(v)
    \quad\Longleftrightarrow\quad
    |\xi_j(x)-\xi_j(v)|\leq\delta_j,
\end{equation}
where $\delta_j$ is chosen relative to the empirical variability of the corresponding model output. 
More precisely,
\begin{equation}
\label{eq:output_tolerance}
    \delta_j
    =
    \operatorname{median}_{x\in\mathcal{X}}
    \left(
        \operatorname{median}_{z\in\mathcal{B}}
        |\xi_j(x)-\xi_j(z)|
    \right),
\end{equation}
with $\mathcal{X}$ denoting the set of explained instances. 
Thus, $\delta_j$ represents a robust estimate of the typical absolute difference between predictions for explained and background instances.
The tolerance is chosen separately for each output component in order to account for differences in scale and empirical variability across the calibrated parameters.

The sample-based $\nu$SHAP values are approximated using the permutation-sampling estimator in \eqref{eq:nushap_perm}. 
Sampling is performed with an estimation-error tolerance $\varepsilon_{\mathrm{MC}}=0.0015$ at a confidence level of $95\%$, with the number of sampled permutations capped at $3\cdot10^5$.

The resulting $\nu$SHAP explanations depend on the selected background dataset and on the input- and output-similarity tolerances, which determine the strictness of the sample-based sufficiency criterion.
Moreover, the absence of a counterexample establishes sufficiency only relative to the finite background dataset. 
In particular, if no background instance satisfies $x_S\approx v_S$, the implication defining sample-based sufficiency holds vacuously. 
The resulting attributions should therefore be interpreted in relation to the stated background sample and tolerance choices.


\section{Experimental setup and calibration performance} \label{sec:cal_performance}

\subsection{Data generation and experimental setup}

The neural networks are trained on synthetically generated implied volatility surfaces with known ground-truth parameters. 
Parameter samples are generated using Latin Hypercube Sampling over bounded parameter domains. 
In the Heston model, we follow the parameter setup of \citet{horvath_2021}. 
In the rough Heston model, the parameter domains follow \citet{yuan_brigo_jacquier_pede_2024}, who adopt parameter ranges originally proposed by \citet{romer_2022} on the basis of empirical SPX data.
Synthetic implied volatility surfaces are computed using the procedure described in Section \ref{sec:framework}.

In the Heston model, we generate $2 \cdot 10^5$ implied volatility surfaces. 
Due to the substantially higher computational cost of option pricing in the rough Heston model, the corresponding datasets contain $10^4$ samples per grid configuration. 
The parameter ranges used throughout the experiments are summarized in Tables~\ref{tab:heston_sampling} and~\ref{tab:rough_sampling}.

\begin{table}[H]
\centering
\renewcommand{\arraystretch}{1.2} 
\begin{tabular}{lrrrrr}
\toprule
& $\kappa$ & $\gamma$ & $\rho$ & $v_0$ & $\bar{v}$\\
\midrule
Lower bound & $0.50$ & $0.01$ & $-0.95$ & $0.0001$ & $0.01$\\
Upper bound & $10.0$ & $1.00$ & $-0.10$ & $0.0400$ & $0.20$\\
\bottomrule
\end{tabular}
\caption{Parameter sampling ranges for the Heston model.}
\label{tab:heston_sampling}
\end{table}

\begin{table}[H]
\centering
\renewcommand{\arraystretch}{1.2} 
\begin{tabular}{lrrrrrr}
\toprule
& $\kappa$ & $\gamma$ & $\rho$ & $v_0$ & $\bar{v}$ & $H$\\
\midrule
Lower bound & $0.1206$ & $0.2720$ & $-0.7071$ & $0.0262$ & $0.0721$ & $0.1286$\\
Upper bound & $0.5041$ & $0.3748$ & $-0.5940$ & $0.0778$ & $0.1499$ & $0.1766$\\
\bottomrule
\end{tabular}
\caption{Parameter sampling ranges for the rough Heston model.}
\label{tab:rough_sampling}
\end{table}

Implied volatility surfaces are discretized over fixed moneyness--maturity grids. 
Moneyness is defined as $m=\frac{K}{S_0}$, where \(K\) is the strike and \(S_0\) the initial spot price. 
Throughout the experiments, \(S_0=1\), so that the strike and moneyness grids coincide. Maturities are expressed in years.

In the rough Heston model, we consider both a long-maturity and a short-maturity discretization, the latter employing a denser concentration of maturities near the origin in order to better capture short-term rough volatility effects. 
The corresponding grids are summarized in Table~\ref{tab:grids}.

\begin{table}[H]
\centering
\renewcommand{\arraystretch}{1.2} 
\begin{tabular}{>{\raggedright\arraybackslash}p{3.4cm}rrr}
\toprule
Model & Moneyness & Maturity & Features\\
\midrule
Heston & $11$ points in $[0.50,\,1.50]$ & $8$ points in $[0.10,\,2.00]$ & $88$\\
rough Heston (long) & $9$ points in $[0.60,\,1.40]$ & $6$ points in $[0.60,\,2.00]$ & $54$\\
rough Heston (short) & $11$ points in $[0.50,\,1.50]$ & $10$ points in $[0.02,\,0.50]$ & $110$\\
\bottomrule
\end{tabular}
\caption{Moneyness--maturity discretisations used throughout the experiments.}
\label{tab:grids}
\end{table}

In the Heston model, the data are divided into training, validation, and test sets containing \(68\%\), \(17\%\), and \(15\%\) of the observations, respectively. 
Analogously, for each rough Heston dataset, the corresponding proportions are \(60\%\), \(20\%\), and \(20\%\).

Prior to training, the input data (implied volatility surfaces) is first scaled using MinMax normalization in order to standardize feature ranges and improve optimization stability. 
The scaled inputs are subsequently preprocessed using Zero-phase Component Analysis (ZCA) whitening, as described by \citet{kessy_lewin_strimmer_2018}.
Whitening removes linear correlations between input coordinates and rescales the transformed data to have approximately unit covariance, thereby improving the numerical conditioning of the optimization
problem. 
Compared with PCA whitening, ZCA whitening preserves the orientation of the original feature space and produces a whitened representation that is closest to the original data in a least-squares
sense. 
Nevertheless, each whitened coordinate remains a linear combination of the original implied-volatility features, as discussed in subsection~\ref{subsec:XAI-frame}.

In the MLP architecture, we consider feedforward neural networks with one to three hidden layers, with width ranging from $16$ to $256$ neurons. 
ReLU activation functions are used throughout, together with Glorot normal initialization \cite{glorot_2010}.
In the highway and softmax-parametrized highway architectures, we consider depths ranging from $3$ to $30$ layers with fixed hidden width $64$. 
The transform branches use ELU activation functions with He normal initialization \cite{he_zhang_2015}, while the gating branches employ sigmoid activation functions with Glorot normal initialization. 
In the softmax-parametrized highway architecture, the transform and carry gates are jointly normalized through the softmax formulation introduced in Section~\ref{sect:architectures}. 
Following \citet{srivastava_greff_schmidhuber_2015}, the transform gate biases are initialized negatively in order to favor carry behavior during the initial training phase.

All networks are implemented in \texttt{TensorFlow} and trained using the \texttt{Adam} optimizer with mean squared error (MSE) loss. 
Model selection is performed using the validation mean absolute error (MAE), which provides a more interpretable measure of calibration accuracy in the present setting. 
Training is performed for $200$ epochs using a batch size of $256$.


\subsection{Calibration performance}

\begin{table}[h]
\centering
\footnotesize
\setlength{\tabcolsep}{3pt}
\renewcommand{\arraystretch}{1.4}

\begin{tabular}{ccc|ccccccc}
\toprule
\textbf{Model} 
& \textbf{Architecture}
& \textbf{Parameters} 
& $\kappa$ 
& $\gamma$ 
& $\rho$ 
& $v_0$ 
& $\bar v$ 
& $H$ 
& \textbf{Avg. MAE}\\
\midrule

\multirow{3}{*}{Heston}
& MLP ($2\times256$)
& 89,861
& 0.0582
& 0.0089
& 0.0069
& $2.4\mathrm{e}{-4}$
& $8.5\mathrm{e}{-4}$
& -- 
& 0.0150
\\

& HW ($7\times64$)
& 67,269
& 0.0346
& 0.0043
& 0.0058
& $1.1\mathrm{e}{-4}$
& $5.2\mathrm{e}{-4}$
& -- 
& 0.0091
\\

& spHW ($15\times64$)
& 197,765
& 0.0307
& 0.0032
& 0.0046
& $2.1\mathrm{e}{-4}$
& $5.8\mathrm{e}{-4}$
& -- 
& $\mathbf{0.0079}$
\\

\midrule

\multirow{3}{*}{
\begin{tabular}{c}
rough Heston \\
(long)
\end{tabular}
}
& MLP ($3\times16$)
& 1,526
& 0.0026
& $9.2\mathrm{e}{-4}$
& 0.0010
& $3.7\mathrm{e}{-4}$
& $6.8\mathrm{e}{-4}$
& $4.2\mathrm{e}{-4}$
& 0.0010 
\\

& HW ($3\times64$)
& 27,526
& 0.0064
& 0.0015
& 0.0018
& $8.7\mathrm{e}{-4}$
& 0.0014
& $7.7\mathrm{e}{-4}$
& 0.0021
\\

& spHW ($30\times64$)
& 376,326
& 0.0011
& $4.5\mathrm{e}{-4}$
& $3.7\mathrm{e}{-4}$
& $2.0\mathrm{e}{-4}$
& $3.6\mathrm{e}{-4}$
& $2.1\mathrm{e}{-4}$
& $\mathbf{4.6\mathrm{e}{-4}}$
\\

\midrule

\multirow{3}{*}{
\begin{tabular}{c}
rough Heston \\
(short)
\end{tabular}
}
& MLP ($1\times16$)
& 1,878
& 0.0096
& 0.0013
& 0.0024
& 0.0016
& 0.0032
& 0.0011
& 0.0032
\\

& HW ($7\times64$)
& 71,558
& 0.0088
& 0.0033
& 0.0035
& 0.0018
& 0.0040
& 0.0018
& 0.0039
\\

& spHW ($30\times64$)
& 390,662
& 0.0084
& 0.0024
& 0.0022
& $8.6\mathrm{e}{-4}$
& 0.0015
& 0.0011
& $\mathbf{0.0027}$
\\

\bottomrule
\end{tabular}
\vspace{0.2cm}
\caption{Calibration performance of the considered neural network architectures. For each model, the best-performing configuration of each architecture is reported together with the corresponding parameter count and per-parameter mean absolute errors.}
\label{tab:calibration_results}
\end{table}

\begin{table}[htbp]
\centering
\footnotesize
\setlength{\tabcolsep}{5pt}
\renewcommand{\arraystretch}{1.35}

\begin{tabular}{cccc}
\toprule
\textbf{Model}
& \textbf{Architecture}
& \multicolumn{2}{c}{\textbf{IV Surface RMSE (vol. points)}} \\
\cmidrule(lr){3-4}
&
& \textbf{Mean}
& \textbf{Median} \\
\midrule

\multirow{3}{*}{Heston}
& MLP ($2\times256$)
& $0.79$
& $0.25$ \\

& HW ($7\times64$)
& $0.81$
& $0.23$ \\

& spHW ($15\times64$)
& $\mathbf{0.74}$
& $\mathbf{0.19}$ \\

\midrule

\multirow{3}{*}{\shortstack{rough Heston\\(long)}}
& MLP ($3\times16$)
& $0.067$
& $0.047$ \\

& HW ($3\times64$)
& $0.17$
& $0.065$ \\

& spHW ($30\times64$)
& $\mathbf{0.041}$
& $\mathbf{0.031}$ \\

\midrule

\multirow{3}{*}{\shortstack{rough Heston\\(short)}}
& MLP ($1\times16$)
& $\mathbf{1.20}$
& $\mathbf{0.15}$ \\

& HW ($7\times64$)
& $1.30$
& $0.17$ \\

& spHW ($30\times64$)
& $2.40$
& $0.18$ \\

\bottomrule
\end{tabular}
\vspace{0.2cm}
\caption{Implied-volatility surface reconstruction errors of the considered
neural network architectures. The mean and median RMSE are computed between
the original implied volatility surfaces and the surfaces repriced using the
predicted parameter sets. Errors are reported in volatility points, where one volatility point corresponds to one percentage point of implied volatility.}
\label{tab:calibration_iv_rmse}
\end{table}

The calibration results for the neural network architectures considered are summarized in Table~\ref{tab:calibration_results}. 
Overall, the networks achieve low test errors within the sampled parameter domains, indicating that they can learn accurate approximations of the inverse map on the synthetic datasets considered here.

In the Heston model, the softmax-parametrized highway architecture achieves the strongest overall performance, followed closely by the highway network. 
Nevertheless, relatively shallow MLP architectures already yield competitive calibration accuracy despite substantially lower architectural complexity. 
Similar behavior is observed for the rough Heston experiments. 
While deeper classical highway architectures tend to exhibit deteriorating performance as depth increases, the softmax-parametrized highway architecture avoids the degradation observed for the classical highway network and generally benefits from increased depth, particularly for the long-maturity rough Heston grid. 
At the same time, these improvements are accompanied by a substantial increase in the number of trainable parameters. 
Across the considered experiments, training and validation errors remained closely aligned for the selected models, suggesting that the reported architectures did not exhibit substantial overfitting.

To complement the parameter-space calibration errors, Table \ref{tab:calibration_iv_rmse} reports the implied-volatility surface reconstruction errors obtained by repricing the surfaces using the predicted parameter sets. For the Heston and rough Heston long-maturity experiments, these results are broadly consistent with the parameter-space errors in Table \ref{tab:calibration_results}, with the softmax-parametrized highway architecture achieving the lowest reconstruction errors. For the rough Heston short-maturity grid, the median reconstruction errors remain relatively small across all architectures, while the substantially larger mean errors indicate a more skewed error distribution with occasional poorly reconstructed surfaces.

The rough Heston short-maturity grid exhibits comparatively higher calibration errors across all architectures. 
This is consistent with the increased complexity of short-maturity rough volatility behavior, which is known to induce highly non-linear implied volatility dynamics. 
Nevertheless, all architectures remain capable of recovering the model parameters with reasonable accuracy. 
In the following section, we analyze the learned inverse mappings using the complementary SHAP and $\nu$SHAP framework.


\section{Structural Analysis results}
\label{sec:results}


\subsection{Global attribution structure}

\begin{figure}[htbp]
    \centering
    \includegraphics[width=\linewidth]{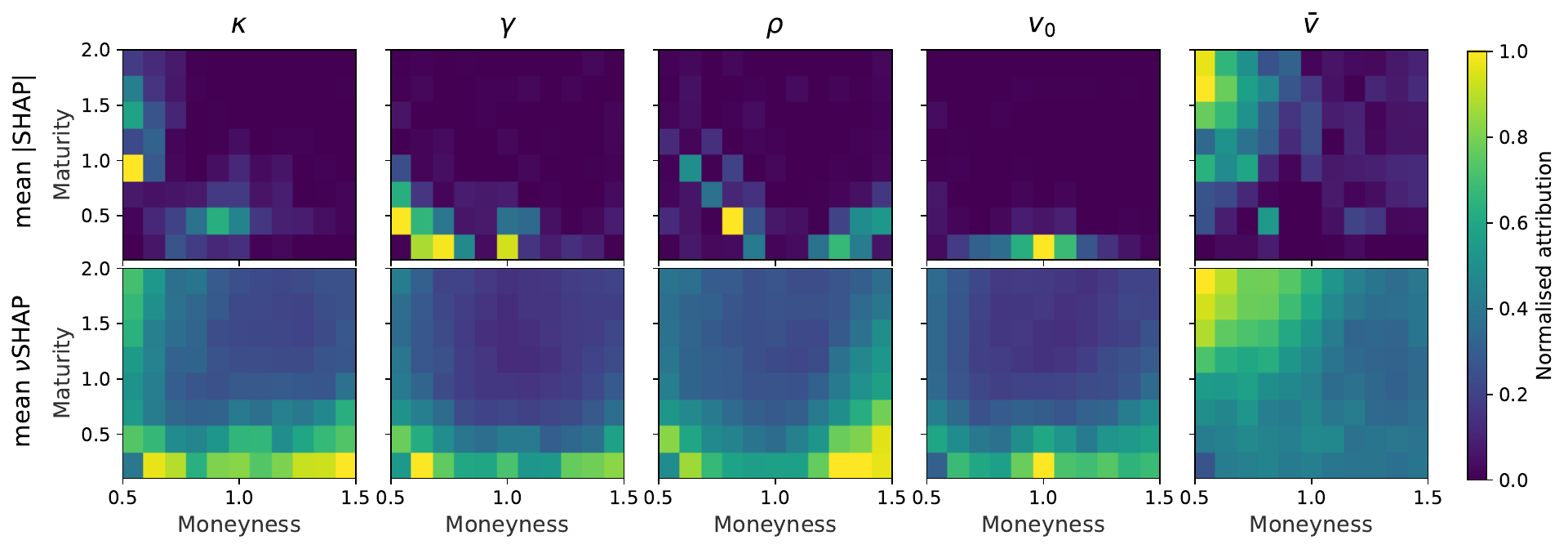}
    \vspace{-0.3cm}
    \caption{Global attribution heatmaps for the Heston model obtained from aggregated SHAP (above) and $\nu$SHAP (below) explanations using the softmax-parametrized highway architecture.}
    \label{fig:heston_attributions}
\end{figure}

\begin{figure}[htbp]
    \centering
    \includegraphics[width=\linewidth]{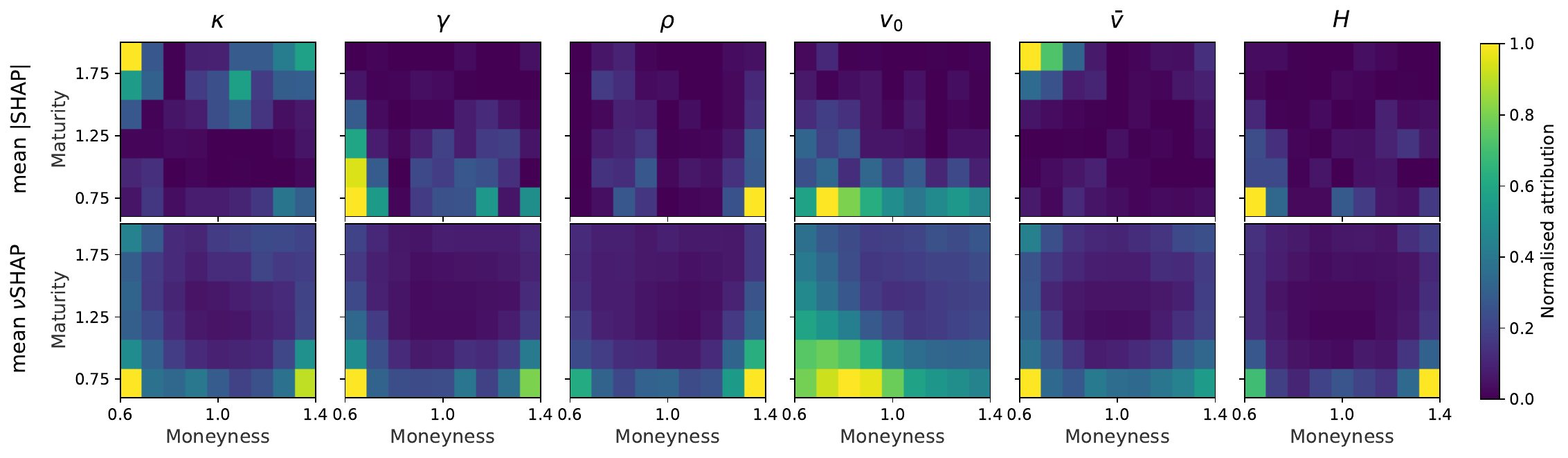}
    \vspace{-0.3cm}
    \caption{Global attribution heatmaps for the rough Heston model on the long-maturity grid obtained from aggregated SHAP (above) and $\nu$SHAP (below) explanations using the softmax-parametrized highway architecture.}
    \label{fig:rough_long_attributions}
\end{figure}

Figures~\ref{fig:heston_attributions} and~\ref{fig:rough_long_attributions} show the normalized global attribution maps for the Heston and the rough Heston models, respectively. 
The attribution scores are arranged according to the dimensions of the original moneyness--maturity grid. 
Since the explanations are computed in the ZCA-whitened feature space, however, the maps represent attribution patterns of transformed coordinates and should not be interpreted pointwise as importance scores for individual option contracts. 

Across both models, the largest relative attribution values tend to occur in coordinates associated with shorter maturities and smile-wing regions. 
The concentration of attribution at shorter maturities is consistent with the theoretical calibration analysis of \citet{forde_jacquier_lee_2012}, which demonstrates how the short-maturity smile encodes information about  -volatility parameters, while variation across maturities contributes to the identification of term-structure parameters such as the mean-reversion rate. In the rough-volatility setting, \citet{fukasawa_2016} establishes a direct connection between the short-maturity implied-volatility skew and the roughness parameter $H$. 

A clear distinction is observed between SHAP and $\nu$SHAP. 
SHAP attributions are generally more localized, highlighting regions where conditioning on individual features strongly changes the expected prediction. 
By contrast, $\nu$SHAP attributions are generally more diffuse and assign importance to broader regions that are sufficient to preserve the predicted parameter value. 
This difference is consistent with the sensitivity--sufficiency distinction discussed in Section~\ref{sec:XAI}. 
Instead of indicating disagreement between the methods, the different attribution structures provide complementary views of the learned calibration map.

\begin{figure}
    \centering
    \includegraphics[width=\linewidth]{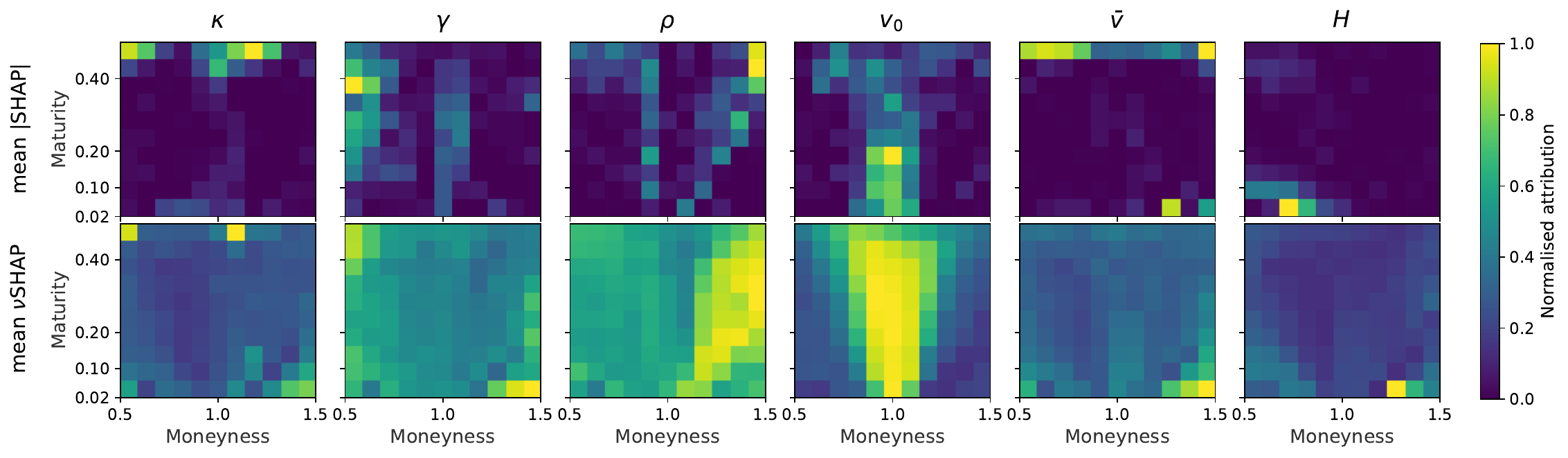}
    \vspace{-0.3cm}
    \caption{Global attribution heatmaps for the rough Heston model on the short-maturity grid obtained from aggregated SHAP and $\nu$SHAP explanations using the softmax-parametrized highway architecture.}
    \label{fig:rough_short_attributions}
\end{figure}

In the rough Heston short-maturity grid, shown in Figure~\ref{fig:rough_short_attributions}, the attribution structure differs noticeably from the long-maturity setting in Figure~\ref{fig:rough_long_attributions}. 
Since the entire grid is concentrated on the short-maturity regime, attribution is distributed more broadly across maturities rather than being concentrated exclusively near the shortest expiries. Nevertheless, both SHAP and $\nu$SHAP continue to assign importance primarily within the short-term region of the implied volatility surface, consistent with the role of short-maturity behavior in rough volatility models. 
While SHAP explanations remain comparatively localized, the corresponding $\nu$SHAP explanations exhibit more diffuse sufficiency patterns, indicating that multiple regions of the short-maturity surface contain overlapping information relevant for parameter recovery.

\subsection{Attribution comparison across architectures}

We next investigate the extent to which the learned attribution structures depend on the chosen neural network architecture. 
While the considered architectures exhibit noticeable differences in predictive calibration performance, it remains unclear whether these differences correspond to fundamentally different inverse mappings, or whether the networks ultimately rely on similar regions of the implied volatility surface.

\begin{figure}[htbp]
    \centering
    \includegraphics[width=\linewidth]{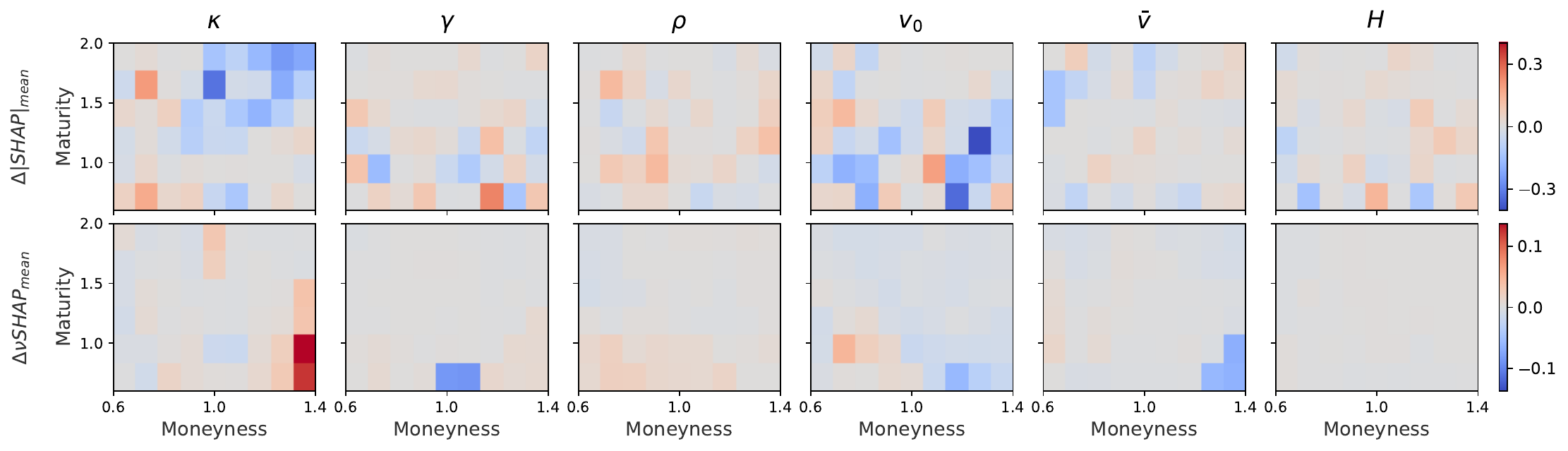}
    \vspace{-0.3cm}
    \caption{Difference heatmaps between the attribution structures of the softmax-parametrized highway and MLP architectures for the rough Heston long-maturity grid.}
    \label{fig:architecture_differences}
\end{figure}

In order to analyze this question, we compare the global attribution structures obtained from the MLP and softmax-parametrized highway architectures for the rough Heston long-maturity grid.
Figure~\ref{fig:architecture_differences} shows the corresponding difference heatmaps for both SHAP and $\nu$SHAP, computed by subtracting the normalized attribution maps of the MLP architecture from those of the softmax-parametrized highway architecture. 
Since the heatmaps are normalized prior to subtraction, the resulting figures reflect relative differences in attribution allocation rather than differences in absolute attribution magnitude.

Several localized differences are visible across both the SHAP and $\nu$SHAP heatmaps. For example, for $v_0$, the SHAP difference map shows a relatively pronounced negative region at short maturities and high moneyness, indicating that the MLP assigns more relative attribution to this part of the surface than the softmax-parametrized highway architecture. By contrast, the $\nu$SHAP difference map for $\kappa$ is nearly uniform across most of the grid, with only a localized deviation in the high-moneyness region. These examples illustrate that, although the architectures redistribute attribution locally, the dominant qualitative attribution structures remain broadly similar. 


This observation is particularly relevant in the context of calibration. 
Although different neural network architectures may realize the inverse map through substantially different parameterizations, the resulting attribution structures suggest that the networks ultimately rely on similar information contained within the implied volatility surface. 
The observed qualitative consistency suggests that the extracted patterns are not entirely specific to one network architecture. 
Nevertheless, this comparison does not constitute a general robustness guarantee, since it is based on the selected MLP and softmax-parametrized highway configurations.


\subsection{Parameter-specific attribution patterns and redundancy}

The attribution structures may be interpreted in light of the known qualitative behavior of stochastic volatility models. 
Prior to analyzing the explanations themselves, it is useful to consider which regions of the implied volatility surface are theoretically expected to contain information regarding the individual model parameters.

The initial variance parameter $v_0$ primarily affects short maturities, since it determines the instantaneous variance level at time $t=0$. 
By contrast, the long-term variance parameter $\bar v$ and the mean-reversion parameter $\kappa$ are expected to exert stronger influence over relatively longer maturities through their effect on long-term variance dynamics and the term structure of implied volatility. 
The volatility-of-volatility parameter $\gamma$ primarily affects the curvature of the smile and is therefore expected to influence the smile wings, particularly at shorter maturities. 
Similarly, the correlation parameter $\rho$ governs the asymmetry of the implied volatility surface and is therefore expected to exhibit attribution concentration away from the at-the-money region. 
In the rough Heston model, the Hurst parameter $H$ primarily affects the steepness of the short-maturity implied volatility skew, particularly within the low-strike wing of the surface.

The observed attribution structures show several qualitative similarities with these expectations. 
In the rough Heston long-maturity grid, both SHAP and $\nu$SHAP explanations frequently assign substantial attribution to the shortest maturities contained within the grid. 
This behavior is consistent with the fact that rough volatility effects manifest most strongly in the short-maturity region of the implied volatility surface, where rough volatility models are known to reproduce steep short-term skew structures; see \citet{gatheral_jaisson_rosenbaum_2018}. 
By contrast, the rough Heston short-maturity grid exhibits attribution patterns that are distributed more broadly across maturities, which is expected since the grid itself is concentrated entirely within the short-term regime. 
Several parameters furthermore exhibit clear attribution concentration within the smile wings, particularly $\rho$ and $\gamma$, whose influence is closely related to skewness and smile curvature.

\begin{figure}[htbp]
    \centering
    \includegraphics[width=0.6\linewidth]{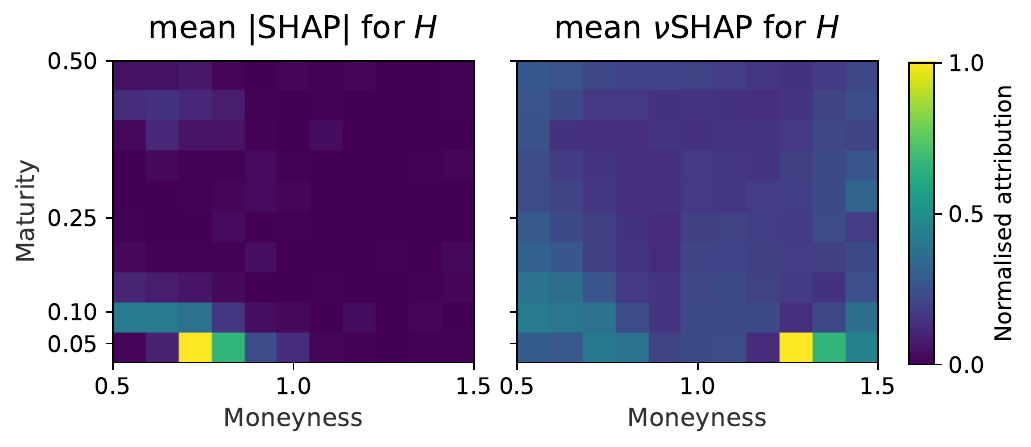}
    \caption{Comparison of SHAP and $\nu$SHAP attribution structures for the Hurst parameter $H$ on the rough Heston short-maturity grid.}
    \label{fig:heatmaps_Hurst}
\end{figure}

Concurrently, the explanations reveal substantial redundancy within the implied volatility surface. 
SHAP explanations often identify relatively localized regions of strong predictive sensitivity, whereas the corresponding $\nu$SHAP explanations are typically more diffuse and distribute attribution across broader neighboring regions. 
This suggests that the predictions do not rely on isolated features, but rather on partially interchangeable subsets of the surface containing overlapping information regarding the model parameters.
Figure~\ref{fig:heatmaps_Hurst} illustrates this distinction for the Hurst parameter $H$ on the rough Heston short-maturity grid. 
While SHAP concentrates attribution within a highly localized region, the corresponding $\nu$SHAP explanation distributes attribution more broadly across the short-maturity smile, indicating that sufficient information for preserving the prediction is distributed across multiple neighboring regions.

Taken together, the attribution patterns provide qualitative evidence that the learned inverse mappings use information broadly consistent with known properties of stochastic volatility models. 


\subsection{Shapley-guided input reduction}
\label{sec:reduction}

The diffuse attribution structures produced by $\nu$SHAP suggest that substantial redundancy is present within implied volatility surfaces. 
In particular, short maturity regions appear to contain sufficient information for preserving the predicted parameter values. 
This raises the question of whether accurate inverse calibration truly requires the full implied volatility surface.

\begin{figure}[htbp]
    \centering
    \includegraphics[width=\linewidth]{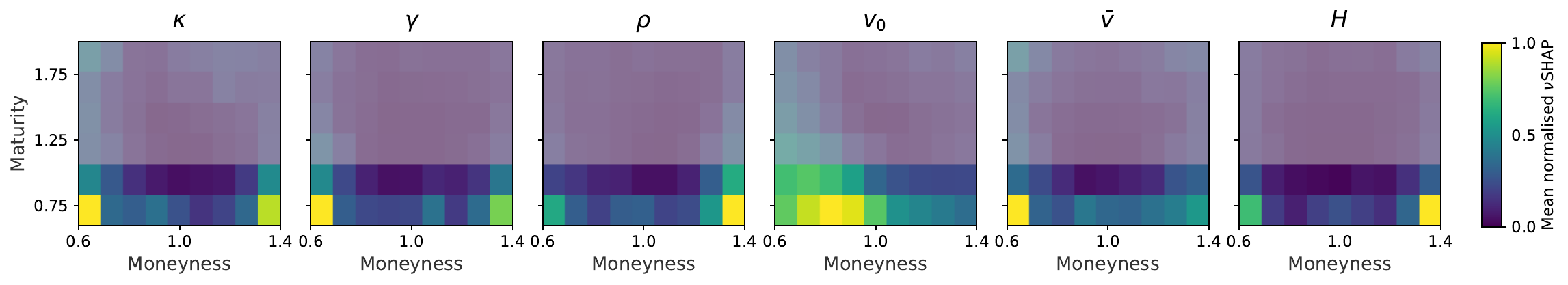}
    \vspace{-0.15cm}
    \caption{Reduced rough Heston long-maturity grid constructed using the global $\nu$SHAP attributions. Greyed regions correspond to discarded features, while the retained short-maturity region contains the subset of the implied volatility surface used for retraining the calibration networks.}
    \label{fig:grid_reduction}
\end{figure}

In order to investigate this, we construct a reduced rough Heston long-maturity grid by retaining only the short-maturity region identified as most relevant by the global $\nu$SHAP attribution maps. 
The retained region is visualized in Figure~\ref{fig:grid_reduction}. 
The reduced grid corresponds to an approximate $67\%$ reduction in input dimensionality, while all remaining regions are discarded.

Using this reduced grid, the softmax-parametrized highway configuration with the lowest MAE for the long-maturity grid was retrained and evaluated using the same setup described in Section~\ref{sec:cal_performance}. 
Table~\ref{tab:grid_reduction} compares the resulting calibration performance obtained using the original and the reduced grid. 
Despite the substantial reduction in input dimensionality, the resulting calibration performance remains comparable to that obtained using the full grid. 
The reduced-grid model even achieves a slightly lower average MAE than the full-grid model. A possible explanation is that the longer-maturity inputs provide only limited additional information for parameter recovery beyond that already contained in the relatively shorter maturities retained in the reduced grid. Removing these redundant or weakly informative inputs may therefore simplify the learning problem and improve generalization, although the observed improvement may also partly reflect training variability.

\begin{table}[h]
\footnotesize
\centering
\renewcommand{\arraystretch}{1.2} 
\begin{tabular}{lrrrrrrr}
\toprule
\textbf{Grid Type} & $\kappa$ & $\gamma$ & $\rho$ & $v_0$ & $\bar{v}$ & $H$ & \textbf{Avg. MAE} \\
\midrule
Full Grid & 0.0011 & $4.5\mathrm{e}{-4}$ & $3.7\mathrm{e}{-4}$
& $2.0\mathrm{e}{-4}$ & $3.6\mathrm{e}{-4}$ & $2.1\mathrm{e}{-4}$ & $4.6\mathrm{e}{-4}$\\
Reduced Grid & $0.0011$ & $3.0\mathrm{e}{-4}$ & $3.0\mathrm{e}{-4}$ & $1.2\mathrm{e}{-4}$ & $3.6\mathrm{e}{-4}$ & $1.7\mathrm{e}{-4}$ & $\mathbf{3.9\mathrm{e}{-4}}$\\
\bottomrule
\end{tabular}
\vspace{0.2cm}
\caption{Calibration performance comparison between the original and $\nu$SHAP-guided reduced rough Heston long-maturity grids using the softmax-parametrized highway architecture.}
\label{tab:grid_reduction}
\end{table}
\begin{table}[h]
\footnotesize
\centering
\renewcommand{\arraystretch}{1.2} 
\begin{tabular}{lcc}
\toprule
\textbf{Grid Type}
& \multicolumn{2}{c}{\textbf{IV Surface RMSE (vol. points)}} \\
\cmidrule(lr){2-3}
& \textbf{Mean}
& \textbf{Median} \\
\midrule
Full Grid & $0.041$ & $0.031$\\
Reduced Grid & $\mathbf{0.026}$ & $\mathbf{0.022}$\\
\bottomrule
\end{tabular}
\vspace{0.2cm}
\caption{Implied volatility surface reconstruction errors for the original and $\nu$SHAP\-guided reduced rough Heston long maturity grids using the softmax-parametrized highway architecture.}
\label{tab:grid_reduction_rmse}
\end{table}

Table \ref{tab:grid_reduction_rmse} provides a complementary evaluation in implied volatility space. 
Consistent with the parameter space results in Table \ref{tab:grid_reduction}, the reduced-grid model also yields lower surface reconstruction errors, with both the mean and median RMSE decreasing relative to the full-grid model. 
As for the improvement in parameter MAE, this may reflect the removal of redundant or weakly informative longer-maturity inputs, although part of the difference may also be attributable to training variability. 

These results provide further evidence that implied volatility surfaces contain substantial redundancy with respect to inverse calibration. 
Rather than relying uniformly on the entire surface, the learned inverse mappings appear to depend primarily on a relatively small subset of highly informative regions. 


\section{Conclusion}
\label{sec:conclusion}

We have investigated the neural network-based inverse calibration of stochastic volatility models from a Shapley-based explainable AI perspective. 
Across the considered experiments, the neural calibration networks achieved low predictive errors for both the Heston and rough Heston models. 
Among the evaluated architectures, the softmax-parametrized highway network generally achieved the best overall calibration accuracy, particularly for the rough Heston experiments. 
At the same time, relatively shallow MLPs produced competitive results with substantially fewer trainable parameters. 
These results demonstrate the practical viability of neural inverse calibration, but predictive accuracy alone does not establish whether the learned mappings are structurally meaningful.

The attribution structures resulting from the XAI analysis were found to align broadly with known qualitative properties of stochastic volatility models. 
In particular, attributions concentrated strongly within short maturities and smile-wing regions known to contain substantial information regarding correlation and volatility-of-volatility parameters, indicating that the networks indeed relied on model-consistent structure.

A central contribution of this paper is the complementary application of the SHAP and $\nu$SHAP explanation methodologies. 
While SHAP explanations frequently identified relatively localized regions of predictive sensitivity, $\nu$SHAP explanations were generally more diffuse and revealed that sufficient information for preserving predictions is often distributed across broader neighboring regions of the surface. 
Rather than viewing these differences as contradictory, the results demonstrated that SHAP and $\nu$SHAP provide complementary perspectives on parameter identifiability and redundancy within implied volatility surfaces. 
The extracted attribution structures were furthermore shown to provide practical guidance for dimensionality reduction. 
Using a reduced rough Heston grid constructed from the global $\nu$SHAP explanations, calibration performance remained comparable to that obtained using the full implied volatility surface despite a substantial reduction in input dimensionality. 
This suggests that Shapley-based methods may serve not only as interpretability tools, but also as practical instruments for feature selection in neural calibration problems.

Several directions remain open for future research. 
First, our analysis was conducted on synthetically generated implied volatility surfaces from the Heston and rough Heston models. 
Extending the proposed structural validation framework to real market data therefore forms an important next step, particularly in the presence of market noise and liquidity effects. 
Furthermore, while we demonstrated that $\nu$SHAP can identify redundant regions of the implied volatility surface and enable substantial input reduction without materially degrading calibration accuracy, future research may investigate more systematic Shapley-guided grid construction and adaptive surface selection procedures. 
Finally, the analysis was restricted to a limited set of neural architectures and stochastic volatility models. 
Future work may therefore study whether the observed attribution structures remain stable across broader model classes, alternative calibration frameworks, and more sophisticated neural architectures.

Overall, the results demonstrate that Shapley-based explainable AI methods can provide useful qualitative insight into neural inverse calibration networks beyond traditional predictive error metrics alone.
More broadly, this work highlights the potential of such methods to improve the transparency, interpretability, and structural validation of neural calibration methods.


\bibliographystyle{abbrvnat}
\bibliography{references}

@article{HeinHofmann2003,
  author  = {Hein, Torsten and Hofmann, Bernd},
  title   = {On the nature of ill-posedness of an inverse problem arising in option pricing},
  journal = {Inverse Problems},
  volume  = {19},
  number  = {6},
  pages   = {1319--1338},
  year    = {2003}}

@article{liu2019neural,
  author        = {Liu, Shuaiqiang and Borovykh, Anastasia and Grzelak, Lech A. and Oosterlee, Cornelis W.},
  title         = {A Neural Network-Based Framework for Financial Model Calibration},
  journal       = {Journal of Mathematics in Industry},
  volume        = {9},
  pages         = {9},
  year          = {2019}
}

@article{baschetti2024deep,
  author        = {Baschetti, Fabio and Bormetti, Giacomo and Rossi, Pietro},
  title         = {Deep Calibration with Random Grids},
  journal       = {Quantitative Finance},
  volume        = {24},
  number        = {9},
  pages         = {1263--1285},
  year          = {2024}
}

@inproceedings{ribeiro2016why,
  author    = {Ribeiro, Marco Tulio and Singh, Sameer and Guestrin, Carlos},
  title     = {"{W}hy {S}hould {I} {T}rust {Y}ou?": {E}xplaining the {p}redictions of {a}ny {c}lassifier},
  booktitle = {Proceedings of the 22nd {ACM} {SIGKDD} International Conference on Knowledge Discovery and Data Mining},
  pages     = {1135--1144},
  year      = {2016},
  publisher = {ACM}
}

@inproceedings{sundararajan2017axiomatic,
  author    = {Sundararajan, Mukund and Taly, Ankur and Yan, Qiqi},
  title     = {Axiomatic Attribution for Deep Networks},
  booktitle = {Proceedings of the 34th International Conference on Machine Learning},
  volume    = {70},
  pages     = {3319--3328},
  year      = {2017}
}

@incollection{Shapley_1953,
    AUTHOR = {Shapley, L. S.},
     TITLE = {A value for {$n$}-person games},
 BOOKTITLE = {Contributions to the theory of games, vol. 2},
     PAGES = {307--317},
 PUBLISHER = {Princeton Univ. Press, Princeton, NJ},
      YEAR = {1953},
   Editors = {H. W. Kuhn and A. W. Tucker}
}

@Article{risks8040101,
AUTHOR = {Cuchiero, Christa and Khosrawi, Wahid and Teichmann, Josef},
TITLE = {A Generative Adversarial Network Approach to Calibration of Local Stochastic Volatility Models},
JOURNAL = {Risks},
VOLUME = {8},
YEAR = {2020},
NUMBER = {4},
ARTICLE-NUMBER = {101}
}

@article{bayer2025on,
  author    = {Bayer, Christian and Horvath, Blanka and Muguruza, Aitor
               and Stemper, Benjamin and Tomas, Mehdi},
  title     = {On Deep Calibration of (Rough) Stochastic Volatility Models},
  journal   = {The Journal of FinTech},
  year      = {2025},
  volume    = {5},
  number    = {1},
  pages     = {2550005}
}

@article{fourier_transform_valuation,
	year = {2010},
	author = {E. Eberlein and K. Glau and A. Papapantoleon},
	title = {Analysis of {F}ourier Transform Valuation Formulas and Applications},
	journal = {Applied Mathematical Finance},
	volume = {17},
	pages = {211--240}
}

@book{Gatheral_2006,
	author = {J. Gatheral},
	title = {The Volatility Surface: A Practitioner's Guide},
	year = {2006},
	publisher = {Wiley}
}

@inproceedings{janzing2020,
  author    = {D. Janzing and L. Minorics and P. Bl\"obaum},
  title     = {Feature relevance quantification in explainable {AI}: A causal problem},
  booktitle = {Proceedings of the 23rd International Conference on Artificial
               Intelligence and Statistics},
  series    = {PMLR},
  volume    = {108},
  pages     = {2907--2916},
  year      = {2020},
}

@unpublished{chen2020,
  author  = {H. Chen and J. D. Janizek and S. Lundberg and S.-I. Lee},
  title   = {True to the model or true to the data?},
  note = {Preprint, arXiv:2006.16234},
  year    = {2020},
}

@inproceedings{sundararajan2020,
  author    = {M. Sundararajan and A. Najmi},
  title     = {The many {S}hapley values for model explanation},
  booktitle = {Proceedings of the 37th International Conference on Machine Learning},
  series    = {PMLR},
  volume    = {119},
  pages     = {9269--9278},
  year      = {2020},
}

@article{aas2021,
  author  = {K. Aas and M. Jullum and A. L{\o}land},
  title   = {Explaining individual predictions when features are dependent:
             More accurate approximations to {S}hapley values},
  journal = {Artificial Intelligence},
  volume  = {298},
  pages   = {103502},
  year    = {2021}
}

@unpublished{roder_dimitroff_2020,
  author        = {R\"oder, D. and Dimitroff, G.},
  title         = {Volatility Model Calibration with Neural Networks:
                   A Comparison Between Direct and Indirect Methods},
  year          = {2020},
  note        = {Preprint, arXiv:2007.03494},
}

@article{hernandez_2016, 
    title={Model calibration with neural networks},
    journal={Risk}, 
    author={Andres Hernandez}, 
    year={2017},
    month = {June}
    }

@article{horvath_2021, 
    author={Blanka Horvath and Aitor Muguruza and Mehdi Tomas}, 
    title={Deep learning volatility: {A} deep neural network perspective on pricing and calibration in (rough) volatility models}, 
    journal={Quantitative Finance}, 
    volume={21}, 
    number={1}, 
    pages={11-27}, 
    year={2021}
}

@unpublished{sridi_bilokon_2023, title={Applying deep learning to calibrate stochastic volatility models}, author={Abir Sridi and Paul Bilokon}, year={2023}, note={Preprint, arXiv:2309.07843}}

@unpublished{srivastava_greff_schmidhuber_2015,
  title={Highway networks},
  author={Rupesh Kumar Srivastava and Klaus Greff and Jürgen Schmidhuber},
  year={2015},
  note={Preprint, arXiv:1505.00387}
}

@unpublished{yuan_brigo_jacquier_pede_2024,
  title={Deep learning interpretability for rough volatility},
  author={Bo Yuan and Damiano Brigo and Antoine Jacquier and Nicola Pede},
  year={2024},
  note={Preprint, arXiv:2411.19317}
}

@article{brigo_huang_pallavicini_haitz_2025, title={Interpretability in deep learning for finance: {A} case study for the {Heston} model}, volume={2}, journal={Risk Sciences}, author={Damiano Brigo and Xiaoshan Huang and Andrea Pallavicini and Haitz Sáez de Ocáriz Borde}, year={2025}, number={100030} }

@article{lundberg_lee_2017, title={A unified approach to interpreting model predictions}, journal={Proceedings of the 31st International Conference on Neural Information Processing Systems}, year={2017}, pages={4768-4777}, author={Scott M. Lundberg and Su-In Lee}}

@unpublished{marquessilva_huang_letoffe_2025, title={The {Explanation} {Game} -- {Rekindled}}, year={2025}, note={Preprint, arXiv:2501.11429}, author={Joao Marques-Silva and Xuanxiang Huang and Olivier Letoffe}}

@InProceedings{marquessilva_2025,
author="Marques-Silva, Joao",
editor="Margaria, Tiziana
and Steffen, Bernhard",
title="Logic-{Based Explainability}: {Past}, {Present} and {Future}",
booktitle="Leveraging Applications of Formal Methods, Verification and Validation. Software Engineering Methodologies",
year="2025",
publisher="Springer Nature Switzerland",
address="Cham",
pages="181--204"
}

@article{black_scholes_1973, title={The pricing of options and corporate liabilities}, volume={81}, number={3}, journal={Journal of Political Economy}, author={Fischer Black and Myron Scholes}, year={1973}, pages={637–654} }

@article{heston_1993, title={A closed-form solution for options with stochastic volatility with applications to bond and currency options}, volume={6}, number={2}, journal={Review of Financial Studies}, author={Steven L. Heston}, year={1993}, pages={327–343} }

@article{fang_oosterlee_2009, title={A novel pricing method for {European} options based on {Fourier-Cosine} series expansions}, volume={31}, number={2}, journal={SIAM Journal on Scientific Computing}, author={Fang Fang and Cornelis W. Oosterlee}, year={2009}, pages={826–848} }

@article{Jackel_2015, author = {Peter Jäckel}, title = {Let's be rational}, journal = {Wilmott}, volume = {2015}, number = {75}, pages = {40-53} }

@article{gatheral_jaisson_rosenbaum_2018, title={Volatility is rough}, author={Jim Gatheral and Thibault Jaisson and Mathieu Rosenbaum}, year={2018}, journal={Quantitative Finance}, volume={18}, number={6}, pages={933-949} }

@article{jaber_larsson_pulido_2019,
      title={Affine {Volterra} processes}, 
      author={Eduardo {Abi Jaber} and Martin Larsson and Sergio Pulido},
   JOURNAL = {The Annals of Applied Probability},
    VOLUME = {29},
      YEAR = {2019},
    NUMBER = {5},
     PAGES = {3155--3200}
}

@article{el_euch_fukusawa_rosenbaum_2018, title={The microstructural foundations of leverage effect and rough volatility}, volume={22}, journal={Finance and Stochastics}, pages={241-280}, author={Omar {El Euch} and Masaaki Fukasawa and Mathieu Rosenbaum}, year={2018}}

@article{el_euch_rosenbaum_2018, title={The characteristic function of rough {Heston} models}, volume={29}, number={1}, journal={Mathematical Finance}, author={Omar {El Euch} and Mathieu Rosenbaum}, year={2019}, pages={3–38} }

@article{diethelm_ford_freed_2004, title={Detailed {Error Analysis} for a {Fractional Adams Method}}, volume={36}, journal={Numerical Algorithms}, author={Kai Diethelm and Neville J. Ford and Alan D. Freed}, year={2004}, pages={31–52} }

@article{mckay_beckman_connover_1979,
author = {M. D. McKay and R. J. Beckman and W. J. Conover},
 journal = {Technometrics},
 number = {2},
 pages = {239-245},
 title = {A comparison of three methods for selecting values of input variables in the analysis of output from a computer Code},
 volume = {21},
 year = {1979}
}

@article{romer_2022, title={Empirical analysis of rough and classical stochastic volatility models to the SPX and VIX markets}, author={Sigurd Emil Rømer}, journal={Quantitative Finance}, volume={22}, number={10}, pages={1805-1838}, year={2022}}

@article{kessy_lewin_strimmer_2018, title={Optimal {Whitening} and {Decorrelation}}, author={Agnan Kessy and Alex Lewin and Korbinian Strimmer}, journal={The American Statistician}, volume={72}, number={4}, pages={309-314}, year={2018}}

@article{glorot_2010,
  title = 	 {Understanding the difficulty of training deep feedforward neural networks},
  author = 	 {Xavier Glorot and Yoshua Bengio},
  journal = 	 {Proceedings of the Thirteenth International Conference on Artificial Intelligence and Statistics},
  pages = 	 {249-256},
  year = 	 {2010},
  volume = 	 {9}
}

@unpublished{he_zhang_2015,
      title={Delving deep into rectifiers: {Surpassing} human-level performance on {Image}{Net} classification}, 
      author={Kaiming He and Xiangyu Zhang and Shaoqing Ren and Jian Sun},
      year={2015},
      note={Preprint, arXiv:1502.01852}
}

@article{forde_jacquier_lee_2012,
author = {Forde, Martin and Jacquier, Antoine and Lee, Roger},
title = {The Small-Time Smile and Term Structure of Implied Volatility under the {H}eston Model},
journal = {SIAM Journal on Financial Mathematics},
volume = {3},
number = {1},
pages = {690-708},
year = {2012}
}

@article{fukasawa_2016, title={Short-time at-the-money skew and rough fractional volatility}, volume={17}, number={2}, journal={Quantitative Finance}, author={Fukasawa, Masaaki}, year={2017}, pages={189–198} }


\end{document}